\documentclass{PoS}

\newcommand{\beq}{\begin{equation}}
\newcommand{\eeq}{\end{equation}}
\newcommand{\bea}{\begin{eqnarray}}
\newcommand{\eea}{\end{eqnarray}}
\newcommand{\D}{\displaystyle} 
\usepackage{epsfig}
\input epsf

\title{Deriving confinement via RG decimations}

\ShortTitle{Deriving confinement via RG decimations}

\author{\speaker{E.T. Tomboulis}\\
        Univ of California, Los Angeles\\
        E-mail: \email{tombouli@physics.ucla.edu}}

\abstract{We present the general framework and building blocks of a recent 
derivation of the fact that the $SU(2)$ LGT is in a confining phase 
for all values of the coupling $0 < \beta < \infty$, 
for space-time dimension $d \leq 4$. The method employs approximate but 
explicitly computable RG decimations that are shown to constrain the exact 
partition function and order parameters from above and below, and flow 
from the weak to the strong coupling regime without encountering 
a fixed point.}

\FullConference{The XXV International Symposium on Lattice Field Theory\\
		 July 30 - August 4 2007\\
		 Regensburg, Germany}

\begin{document}

\section{Introduction}  

4-dim. $SU(N)$ gauge theory at $T=0$ is known to be in a single phase 
for all values of the gauge coupling, $ 0 < \beta <\infty$. 
This fact has proved hard to derive directly. This is not surprising since 
one is faced with a multi-scale problem involving passage  
from a short distance perturbative ordered regime to a long distance 
non-perturbative confining disordered regime. The natural framework  here  
is RG block-spinning bridging these disparate regimes.     
Ideally, one would like to construct an exact block-spinning scheme 
converging to the `perfect action' governing the Wilsonian 
renormalized trajectory. Then one would be able to compute any observable 
at different scales. Despite some valiant efforts, however, this has proven 
technically too complicated to carry out so far.

Progress can be made, nonetheless, by adopting a more modest but 
still sufficiently general framework. This approach 
employs approximate, rather than exact, 
but easily explicitly computable decimation 
procedures that can provide {\it bounds} on judicially chosen quantities. 
Such quantities are partition functions (free energies) and   
their ratios (free energy differences). 
The basic idea is to consider only such quantities - they can serve as order 
parameters - and not attempt to construct a general RG blocking action 
suitable, in principle, for computing any observable.  
One then uses the bounds to fix the behavior of corresponding exact 
quantities by interpolating  between the bounds. In this fashion one 
can derive statements concerning the exact theory, and in particular 
the presence of the confining phase. A detailed account has recently 
appeared in \cite{TT}.  
Here we present an outline of the basic steps involved in this development. 

\section{Partition function}  
We employ standard lattice gauge theory (LGT) notations, $U$ denoting 
generic group elements, 
$U_b$ bond variables, $U_p=\prod_{b\in \partial p}U_b$, etc.  
Starting  with a plaquette action, for example the Wilson 
action 
\beq 
A_p(U_p) ={\beta\over 2}\;\chi_{1/2}(U_p), 
\eeq 
the character expansion of the exponential of the action is:
\beq 
e^{A_p(U)} 
   = \sum_j\;d_j\,F_j(\beta)\,\chi_j(U) \,.  
\eeq 
For $SU(2)$, the only case considered explicitly here, $j=0, {1\over 2}, 1, 
{3\over 2}, \ldots$, and $d_j=(2j+1)$. 
In terms of normalized coefficients $c_j(\beta) = F_j(\beta)/F_0(\beta)$: 
\beq
e^{A_p(U)} = F_0\,\Big[\, 1 + \sum_{j\not= 0} d_j\, c_j(\beta)\,
\chi_j(U)\, \Big] \, . \label{exp1}
\eeq
The Partition Function (PF) on lattice $\Lambda$ of spacing $a$ is then 
defined by 
\beq
Z_\Lambda(\beta) = \int dU_\Lambda\;\prod_p\, 
\Big[\, 1 + \sum_{j\not= 0} d_j\, c_j(\beta)\,
\chi_j(U)\,  \Big] \equiv Z_\Lambda\Big(\{c_j(\beta)\}\Big) \label{PF1}
\eeq 
For a reflection positive action:
\beq F_j \geq 0\qquad \mbox{hence}\quad 1\geq c_j\geq 0 \qquad\quad 
\mbox{all}\quad j \;.
\eeq 

We now introduce decimations whereby the lattice spacing is 
changed by a scale factor $b$. We employ  
decimation operations of the 'bond moving type' \cite{M}, \cite{K}, 
which preserve the form (\ref{PF1}). 
Such a decimation operation can be 
summarized as a set of decimation  
rules for each successive step:
\bea 
& & a \to b a \to b^2 a \to \cdots \to b^n a \nonumber \\
 & & \Lambda \to \Lambda^{(1)} \to \Lambda^{(2)} \to \cdots \to 
\Lambda^{(n)} \nonumber
\eea 
from lattice $\Lambda^{(m)}$ of spacing $b^{m}a$ to 
lattice $\Lambda^{(m+1)}$ of spacing $b^{m+1}a$, with  
$\Lambda^{(0)}=\Lambda$.

The rules provide an explicit expression for the computation of the Fourier 
coefficients at the $m+1$-th step given those of the $m$-th step: 
\bea 
F_0(m) & = & F_0(\zeta, r, b, \{ c_i(m-1)\}) \label{RG1} \\
c_j(m) & = & c_j(\zeta, r, b, \{ c_i(m-1)\}) \; . \label{RG2} 
\eea 
The explicit form of (\ref{RG1}) - (\ref{RG2}) need not be given here; we 
only note that they involve parameters $\zeta$, $r$ which control the amount 
by which the interactions of the plaquettes remaining after a decimation 
step are `renormalized' to compensate for the ones that were removed. 

Correspondingly, the partition function undergoes the transformation
\beq
Z_{\Lambda^{(m-1)}}\Big(\{c_j(m-1)\}\Big) \to 
F_0(m)^{|\Lambda^{(m)}|} \, Z_{\Lambda^{(m)}}\Big(\{c_j(m)\}\Big) \;.\label{PF2}
\eeq 
There is a bulk free energy contribution from the blocking $b^{m-1} a \to
b^{m} a$. 
The resulting effective action in $Z_{\Lambda^{(m)}}\Big(\{c_j(m)\}\Big)$ 
on the resulting lattice $\Lambda^{(m)}$ retains the original one-plaquette 
form. It will, however, 
contain, in general, all group representations even after just one 
decimation step 
starting from an action containing one or a finite number of representations:  
\beq
\exp \left[\, A_p(U_p, m)\,\right] = 
\Big[\, 1 + \sum_{j\not= 0} d_j\, c_j(m)\,\chi_j(U_p)\,
 \Big]  \label{exp2}
\eeq 
with   
\beq
A_p(U, m)= \sum_j \; \beta_j(m)\,\chi_j(U) \;.
\eeq
Also, {\it both} positive and negative effective couplings 
$\beta_j(m)$ will in general occur. 
But in (\ref{exp2}) all $c_j(m)\geq 0$ if the parameter 
$\zeta={\rm integer}$. This implies that reflection positivity of the 
measure is preserved after each decimation step. 

\vspace{0.3cm}
\noindent {\it Upper and lower bounds \ } 
In going from the $(m-1)$-th  step to the $m$-th decimation step consider the 
following choices of decimation parameters in (\ref{RG1}) - (\ref{RG2}): 
\begin{itemize}
\item (I) $\zeta=b^{d-2}$, $\quad r=1-\epsilon\;, \qquad 
0\leq \epsilon < 1$; 
denote the resulting $m$-th step coefficients by $F_0^U(m)$ and $c_j^U(m)$. 
This is essentially (for $r=1$) the choice made in \cite{M} - \cite{K}.   
\item (II) $\zeta=1$, $\quad r=1 \;$;    
denote the resulting coefficients by $F_0^L(m)$ and $c_j^L(m)$.  
\end{itemize}
Then one has the following basic inequalities relating the  
PF's before and after the decimation: 
\beq 
F_0^L(m)^{|\Lambda^{(m)}|}\,Z_{\Lambda^{(m)}}\Big(\{c_j^L(m)\}\Big) \, < 
Z_{\Lambda^{(m-1)}} 
< \,F_0^U(m)^{|\Lambda^{(m)}|}\,Z_{\Lambda^{(m)}}\Big(\{c_j^U(m)\}\Big) 
\;. \label{I}
\eeq 
A variety of similar bounds can be employed 
(i.e. somewhat different definitions of $F_0^{U,\,L}(m)$ and $c_j^{U,\,L}(m)$) 
in (\ref{I}) (cf. \cite{TT}). Such technical details are not important for 
the general development below.

\vspace{0.3cm}
\noindent {\it Interpolation \ }
Introducing a parameter $\alpha$, $0\leq \alpha\leq 1$, define  
interpolating coefficients $\tilde{c}_j(m,\alpha)$ and 
$\tilde{F}_0(m,\alpha)$ 
such that 
\beq \tilde{c}_j(m,\alpha)= \left\{ \begin{array}{lll}
c_j^U(m) & : &\alpha=1 \\ 
c_j^L(m)  & : & \alpha=0 
\end{array} \right. \;,
\eeq 
and similarly 
\beq \tilde{F}_0(m,\alpha)= \left\{ \begin{array}{lll}
F_0^U(m) & : &\alpha=1 \\ 
F_0^L(m) & : & \alpha=0 
\end{array} \right. 
\eeq  
There is clearly an infinity of such smooth interpolations that can be 
defined. 
But there is nothing unique about any one such interpolation. It is 
expedient then to consider more generally a family of smooth 
interpolations parametrized 
by a parameter $t$ in some interval  $(t_1, t_2)$.

Then the upper-lower bounds statement (\ref{I})
implies that, for each value of $t$ picking an interpolation family member, 
there exist some value of the interpolating parameter 
$\alpha=\alpha^{(m)}_\Lambda(t)$, where  
\[ 0 < \alpha^{(m)}_\Lambda(t) < 1 \;,\]
such that  
\beq 
\tilde{F}_0(m,\alpha,t)^{|\Lambda^{(m)}|} \,  
Z_{\Lambda^{(m)}}\Big(\{\tilde{c}_j(m,\alpha)\}\Big) 
= Z_{\Lambda^{(m-1)}} \,.
\label{Ifix} 
\eeq
Note that, by construction, there is parametrization invariance under 
shift in t in the l.h.s. of (\ref{Ifix}); in other words,  
$\alpha=\alpha^{(m)}_\Lambda(t)$ is the level surface fixed
by (\ref{Ifix}). Furthermore, one can show that  
\beq \alpha^{(m)}_\Lambda(t)= \alpha^{(m)}(t) + \delta\alpha^{(m)}_\Lambda(t)
\;, \qquad \mbox{with}\qquad   \delta\alpha^{(m)}_\Lambda(t)\to 0, \quad 
|\Lambda|\to \infty  \label{alph}
\eeq
and lattice-volume independent $\alpha^{(m)}(t)$. 

So, iterating this procedure starting from the original lattice, one gets 
an {\it exact integral representation} of the PF 
on successively decimated lattices: 
\bea 
Z_\Lambda(\beta) & = & Z_\Lambda\Big(\{c_j(\beta)\}\Big) 
\nonumber \\
&  = & \tilde{F}_0(1,\alpha_\Lambda^{(1)}(t_1),t_1)^{|\Lambda^{(1)}|} \,  
Z_{\Lambda^{(1)}}\Big(\{\tilde{c}_j(1,\alpha_\Lambda^{(1)}(t_1))\}\Big)
\nonumber \\
& = & \cdots  \nonumber \\
& = & \Big[\prod_{m=1}^n 
\tilde{F}_0(m,\alpha_\Lambda^{(m)}(t_m),t_m)^{|\Lambda|/b^{dm}} \Big]
\,Z_{\Lambda^{(n)}}\Big(\{\tilde{c}_j(n,\alpha_\Lambda^{(n)}(t_n))\}\Big) 
\label{A}
\eea
The representation is in terms of the accumulated bulk free energy 
contributions from the successive blockings from scale $a$ to 
scale $b^n a$, and the resulting effective action and corresponding 
PF on $\Lambda^{(n)}$. Note that all coefficients occurring in 
this representation are constrained by the bounding coefficients that 
are computable by the decimation rules (\ref{RG1}) - (\ref{RG2}). 

There are many potential uses for exact representations such as (\ref{A}) 
and its derivatives. In the following we use it to examine the 
confining properties of the theory. 

\section{`Twisted' partition function}
Let ${\cal V}$ denote a coclosed set of plaquettes winding around the 
periodic lattice in the $d-2$ directions normal to, say, the $[12]$-plane.  
Let $Z_\Lambda^{(-)}\;$ denote the 
partition function with action on every plaquette in ${\cal V}$ shifted by 
a non-trivial  element $\tau$ (`twist') of the group center. 
Thus, for $SU(2)$, 
$\tau=-1 \in Z(2)$. The twist represents the introduction of external 
$\pi_1(SU(2)/Z(2)) = Z(2)$ vortex flux in $\Lambda$. 
For $Z_\Lambda^{(-)}$ reflection positivity holds only in planes 
perpendicular to the directions in which ${\cal V}$ winds around the lattice. 
To have RP in all planes one may simply 
replace $Z_\Lambda^{(-)}$ by $Z^+_\Lambda \equiv 
{1\over 2}\Big(Z_\Lambda + Z_\Lambda^{(-)}
\Big)$.

The above development can then be carried through also for $Z^+_\Lambda$ 
applying the same decimations (\ref{RG1}) - (\ref{RG2}), obtaining 
the analog of (\ref{I}) giving upper and lower bounds in terms of 
$F_0^{U,\,L}(m)$ and $c_j^{U,\,L}(m)$, 
and interpolating between them. One thus obtains the corresponding integral 
representation on successively decimated lattices:  
\bea 
Z^+_\Lambda 
& = & \Big[\prod_{m=1}^n 
\tilde{F}_0(m,\alpha_\Lambda^{+(m)}(t_m),t_m)^{|\Lambda|/b^{dm}} \Big] 
\nonumber\\
& &  \cdot\; {1\over 2} 
\Big[\,Z_{\Lambda^{(n)}}\Big(\{\tilde{c}_j(n,\alpha_\Lambda^{(+)}(t_n))\}\Big) 
+ Z_{\Lambda^{(n)}}^{(-)}\Big(\{\tilde{c}_j(n,\alpha_\Lambda^{+(n)}(t_n))\}
\Big) \Big]  \,.\label{B}
\eea
Again, one can show that  
\beq \alpha^{+(m)}_\Lambda(t)= \alpha^{(m)}(t) + \delta\alpha^{+(m)}_\Lambda(t)
\;, \qquad \mbox{with} \qquad \delta\alpha^{+(m)}_\Lambda(t)\to 0,
\quad|\Lambda|\to \infty \,. \label{alph+}
\eeq
As can be seen from (\ref{B}), the external flux presence does not affect 
the bulk free-energy contributions that resulted from successive blockings. 
Also, it should be noted that, as indicated by the notation,   
the values $\alpha_\Lambda^{+(m)}(t)$ fixed at each successive step 
$m=1,\ldots,n$ in this representation are a priori 
distinct  from the values 
$\alpha_\Lambda^{(m)}(t)$ in the representation (\ref{A}) for  
$Z_\Lambda(\beta)$. This is because they are 
fixed by an independent procedure 
involving a distinct quantity. It is easily seen, however, that any 
discrepancies between $\alpha_\Lambda^{+(m)}(t)$ and 
$\alpha_\Lambda^{(m)}(t)$ can only occur in the lattice-dependent parts 
$\delta\alpha^{+(m)}_\Lambda(t)$, $\delta\alpha^{(m)}_\Lambda(t)$. 
This assumes that the same family of 
interpolations is used in (\ref{A}) and (\ref{B}). 
In general, however, one may of course make different choices of 
interpolation in the two cases.

\section{Order parameters - Vortex free energy} 
The vortex free-energy is defined as: 
\beq
\exp ( - F_\Lambda(\beta)) = {Z_\Lambda^{(-)}\over Z_\Lambda} \;.
\label{vfe1}
\eeq
Physically, it represents the difference in free energies 
between the vacuum in the presence and in the absence of an 
externally introduced $\pi_1(SU(2)/Z(2))$ vortex. As it is well-known,  
(\ref{vfe1}) serves as an order parameter characterizing 
the phases of gauge theory \cite{tH}. It is known, in particular, that 
confining behavior for (\ref{vfe1}) implies confining behavior 
(area law) of the Wilson loop \cite{KT}.  

One may now represent this ratio on successively decimated lattices 
by inserting our representations (\ref{A}), (\ref{B}) 
in the numerator and denominator in 
\beq 
\left( 1 + {\D Z_\Lambda^{(-)}\over \D Z_\Lambda}\right) 
=   {Z_\Lambda + Z_\Lambda^{(-)} \over Z_\Lambda} \,. \label{vfe2}
\eeq
Comparing these representations one sees that any discrepancies 
between $\alpha_\Lambda^{+(m)}(t)$ and 
$\alpha_\Lambda^{(m)}(t)$, even if they vanish in the large volume limit 
(cf. (\ref{alph}), (\ref{alph+})),  
can leave residual non-vanishing effects from bulk free energy 
contributions not completely canceling between numerator and denominator 
in (\ref{vfe2}).  
One may, however, utilize the independent invariance under parametrization 
shifts in numerator and denominator to arrange for complete cancellation 
of the bulk free energy pieces between numerator and denominator 
generated at each step.  

Carrying out  $n$ decimation steps with $n$ sufficiently large one ends up 
with 
\beq
 {\D Z_\Lambda^{(-)}\over \D Z_\Lambda} = 
{Z_{\Lambda^{(n)}}^{(-)}\Big(\{\tilde{c}_j(n,\alpha_\Lambda^{*(n)})\}
\Big)  \over 
Z_{\Lambda^{(n)}}\Big(\{\tilde{c}_j(n,\alpha_\Lambda^{*(n)})\}
\Big)} \;. \label{vfe3}
\eeq
Here $\alpha_\Lambda^{*(n)}$ denotes 
$\alpha_\Lambda^{(n)}(t)$ at a particular $t=t^*$ such that 
$\alpha_\Lambda^{(n)}(t^*)=\alpha_\Lambda^{+(n)}(t^*)$ after a final 
interpolation parametrization shift in the numerator versus the denominator.  

Now, by construction, the coefficients  in terms of 
which the PF's in (\ref{vfe3}) are computed are bounded by the upper bound 
coefficients: 
\beq 
\tilde{c}_j(n,\alpha_\Lambda^{*(n)}) \leq c_j^U(n)  \,. \label{cbound1}
\eeq
Recall that the upper bound coefficients are explicitly computable in terms of 
the decimation rules, (\ref{RG1}), (\ref{RG2}). 
For the type of potential moving decimations used here it is a known fact 
that, for $SU(2)$, one finds 
\beq
c_j^U(n) \to 0\, \qquad 
\mbox{for} \quad\quad n\to \infty \, , \label{cbound2} 
\eeq
provided the space-time dimensionality $d \leq 4$. 
(\ref{cbound1}), (\ref{cbound2}) imply then that, by taking the 
number of decimations $n$ large enough, one can compute the resulting 
expression (\ref{vfe3}) within the 
convergent strong coupling expansion. 

In this manner one arrives at the following: 
\begin{itemize}
\item The vortex free energy parameter (\ref{vfe1}) exhibits 
confining behavior for any 
initial $\beta$ and $d\leq 4$. This follows from the fact that 
(\ref{cbound2}) holds for any initial $\beta$. 
\item Fixing the resulting string tension $\; \kappa (\beta,n)$  
implies a relation between $n$ and $\beta=2/g^2$. 
\item Now zero coupling $g=0$ is a fixed point of the decimations. 
This implies  
that to reach any fixed value of the string tension (some given value of 
$c_j^U(n)$'s) requires 
\beq
\beta \to \infty   \quad \Longleftrightarrow  \quad n \to \infty   \;.
\eeq
In other words one necessarily has  
\beq
 g(a) \to 0 \qquad  \mbox{for} \qquad a\to 0 
\eeq
as an essentially qualitative feature of the decimation flow. 
\end{itemize}

\section{Summary} 

A framework was developed within LGT that utilizes approximate but 
explicitly computable RG decimations which provide bounds constraining 
the exact theory. By a process of interpolation between such lower and 
upper bounds statements concerning the behavior of the exact theory can be 
obtained. This has many potential applications. 

This framework was applied to the $SU(2)$ gauge theory.  
Exact integral representations of the PF with or without 
external flux on successively coarser lattices were obtained. There are 
many potential uses for such representations. 
They were used here to examine the so-called vortex free-energy order 
parameter which can characterize the phases of the theory. 
Confinement at $T=0$ emerges for any initial coupling in four 
space-time dimensions once the approximate 
bounding decimations possess this property.  		

Extension to $SU(3)$ should be straightforward -- the approximate 
decimations exhibit the same qualitative flow as in the $SU(2)$ case. 
In the above  we were able to extract statements about the exact theory 
without actually knowing the actual numerical values of the 
interpolating parameters in the 
exact PF representations. Developing methods for numerical 
approximation of these values would be very useful, for example in 
conjunction with MCRG (Monte Carlo RG) techniques.  

\acknowledgments
This work was partially supported by NSF-PHY-0555693.

\end{document}